\documentstyle[prb,aps,epsf]{revtex}
\begin{document}
\draft
\title{
Marginal Fermi liquid behavior from 2d Coulomb interaction}
\author{J. Gonz\'alez$^{\dag}$, F. Guinea$^{\ddag}$ and
M. A. H. Vozmediano$^*$ \\}
\address{
$^{\dag}$ Instituto de Estructura de la Materia.  Consejo
Superior de Investigaciones Cient{\'\i}ficas.  Serrano 123,
28006-Madrid. Spain. \\ $^{\ddag}$ Instituto de Ciencia de
Materiales.  Consejo Superior de Investigaciones
Cient{\'\i}ficas.  Cantoblanco. 28049 Madrid. Spain.  \\ $^*$
Escuela Polit\'ecnica Superior. Universidad Carlos III.
Butarque 15. Legan\'es. 28913 Madrid. Spain.}
\date{\today}
\maketitle 
 
\begin{abstract}
A full, nonperturbative renormalization group analysis of interacting 
electrons in a graphite layer is performed, in order to 
investigate the deviations from Fermi liquid theory
that have been observed in the experimental measures of a 
linear quasiparticle decay rate in graphite.
The electrons are coupled through
Coulomb interactions, which remain unscreened due to the semimetallic
character of the layer. We show that the model 
flows towards the noninteracting fixed-point
for the whole range of couplings, with logarithmic corrections
which signal the marginal character of the interaction separating
Fermi liquid and non-Fermi liquid regimes.
\end{abstract}



During recent years there has been important progress in
understanding the properties of quantum electron liquids in
dimension $D < 3$. One of the most fruitful approaches in this
respect springs from the use of renormalization group (RG)
methods, in which the different liquids are characterized by
several fixed-points controlling the low-energy properties. The
Landau theory of the Fermi liquid in dimension $D > 1$ can be
taken as a paradigm of the success of this program. It has been
shown that, at least in the continuum limit, a system with
isotropic Fermi surface and regular interactions is susceptible
of developing a fixed-point in which the interaction remains
stable in the infrared\cite{sh}. 

The question of whether different critical points may arise at
dimension $D = 2$ is now a subject of 
debate\cite{a,feld,fk,wm}. From the perspective of the RG approach, 
one of the premises
leading to the Fermi liquid fixed-point should be relaxed in
order to flow to a different universality class. In the case of
models proposed to understand the electronic properties of
copper-oxide superconductors, the high anisotropy of the Fermi
surface\cite{shen}
may play an important role in the anomalous behavior of
the normal as well as of the superconducting state\cite{inst}. 
On the other hand, a possible source of
non-Fermi liquid behavior may arise in systems with singular
interactions\cite{wen}.
In the case of the Coulomb interaction screened by the Fermi sea, 
a solution by means
of bosonization methods has shown that no departures from Fermi 
liquid behavior arise at $D =2$ and 3\cite{shankarb}.
It has been also shown for the conventional screened interaction
that only potentials as singular as $V(q) \sim 1/|q|^{2D - 2}$
can lead to a different electron liquid\cite{wen,ital}. 
The system of electrons with magnetic interactions is quite 
different in that respect. It is known that in this case
the system shows non-Fermi liquid behavior for $D \leq 3$, manifested in 
properties like the specific heat or anomalous electron field
dimensions\cite{wong,wilczek,chakrav}.

In the present work, we address the applicability of the notion of
Fermi liquid fixed-point to two-dimensional systems with
unscreened Coulomb interaction. The absence of screening requires
the vanishing of the density of states at the Fermi level.
Semimetals show this behavior, while retaining a gapless electronic
spectrum. The existence of a well defined continuum at low energies
permits the existence of non trivial scaling properties\cite{us}.
The most
remarkable example of this kind is given by the two-dimensional
sheet of graphite, which has a vanishing density of states at
the Fermi level\cite{us2}. 
Recent photoemission
experiments in graphite, at intermediate
energies, show a decay rate of quasiparticles
proportional to their energy\cite{exp}. 
This represents a clear deviation
with respect to the behavior in metals, which follow the
conventional Fermi liquid picture with quasiparticle lifetimes
propoportional to the inverse of the energy square, with, at most,
logarithmic corrections. A
description in terms of an effective field theory model has
shown that the electronic interactions within the graphite
layers are mainly responsible for the anomalous properties
measured in the experiment\cite{unc}.

We apply RG techniques to investigate whether the
mentioned anomalous behavior can be understood as a marginal
deviation from Fermi liquid theory, or rather it points towards
a different universality class realized in the graphite sheet.
We recall that the low-energy electronic excitations of the
latter at half-filling are concentrated around two Fermi points
at the corners of the hexagonal Brillouin Zone, where the
dispersion relation is well approximated by two
cones in contact at the apex.
The effective field
theory is given therefore by a pair of Dirac fermions, with a
Coulomb potential that remains unscreened due to the vanishing
density of states at the Fermi points. The effective hamiltonian
can be written in the form\cite{foot}
\begin{equation}
H = -i v_F \int d^2 r \Psi^{+}({\bf r}) \mbox{\boldmath $\sigma
\cdot \nabla $} \Psi ({\bf r}) + \frac{e^2}{8 \pi} \int d^2 r_1
\int d^2 r_2 \Psi^{+}({\bf r}_1)\Psi({\bf r}_1) \frac{1}{|{\bf r}_1
- {\bf r}_2|} \Psi^{+}({\bf r}_2) \Psi({\bf r}_2)
\label{ham}
\end{equation}
where $\Psi ({\bf r})$ is a two-dimensional Dirac spinor and $
\mbox{\boldmath $\sigma $} \equiv (\sigma_x , \sigma_y )$. Such
effective field theory provides a good starting point for a RG
analysis since, given that the scaling dimension of the
$\Psi({\bf r})$ field is $-1$ (in length units), the
four-fermion Coulomb interaction turns out to be scale
invariant, at this level, with a dimensionless coupling constant
$e^2$.

In order to address the existence of a new universality class,
besides the trivial noninteracting phase,
a nonperturbative approach has to be
adopted, since such new phase can only be revealed by
a nontrivial fixed-point in coupling constant space.  In the following,
we will implement a GW approximation in the computation
of the self-energy properties. This is more easily achieved in
the present model by replacing the four-fermion term in
(\ref{ham}) by the interaction with an auxiliary scalar field
used to propagate the Coulomb interaction. The effective
hamiltonian can be rewritten in the form
\begin{equation}
H = -i v_F \int d^2 r \Psi^{+}({\bf r}) \mbox{\boldmath $\sigma
\cdot \nabla $} \Psi ({\bf r}) + e \int d^2 r
 \Psi^{+}({\bf r})\Psi({\bf r}) \; \phi ({\bf r})
\end{equation}
where the scalar field $\phi ({\bf r}) $ has the propagator
\begin{equation}
i \langle T  \phi ({\bf r},t) \; \phi ({\bf r}',t') \rangle =
\frac{1}{4\pi} \delta (t - t') \frac{1}{|{\bf r} - {\bf r}'|}
\end{equation}

In this framework we will introduce the GW approximation by
taking into account the quantum corrections to the $\phi $
propagator due to particle-hole excitations of the Fermi sea.
This kind of approximation has proven to be adequate to the
description of the crossover from Fermi liquid to Luttinger
liquid behavior upon lowering the dimension from $D = 2$ to $1$,
capturing the relevant physical processes in the electron
system\cite{metz}. 
Therefore, it seems also appropriate to uncover any
possible fixed-point, different from that of Fermi liquid
theory, in the case of the system with unscreened Coulomb
interaction.

The perturbative analysis of our model shows, in fact, the
existence of a free fixed-point that is stable in the infrared
limit. This can be understood from the nontrivial scaling of the
model with respect to variations of the bandwidth cutoff $E_c$,
that is needed to regulate the divergent contribution of virtual
processes to observable quantities. According to the RG point of
view, a reduction of the cutoff $E_c$ implements a partial
integration of the high-energy electron modes, that renormalize
in this way the value of the effective parameters in the
low-energy theory. The electron charge $e$ is not renormalized
in the present model, but the Fermi velocity $v_F$ is
renormalized to first order in perturbation theory by a
self-energy correction of the form\cite{us}
\begin{equation}
\Sigma ({\bf k}, \omega_k ) \approx  \frac{e^2}{8 \pi}
\mbox{\boldmath $\sigma \cdot k$} \log E_c
\end{equation}
Thus, in the perturbative regime the Fermi velocity $v_F$ grows
steadily as the cutoff $E_c$ is reduced, and the effective 
coupling constant $e^2 / v_F$
flows to zero in the low-energy effective theory.

The most interesting point, however, concerns the analysis of
the model away from the perturbative regime. 
With regard to the direct
application to the graphite system the weak coupling results are
of little use, since the bare coupling in the graphite sheet has
an estimated value $e^2 / v_F \sim 10$. Some possibly
relevant effects like, for instance, the renormalization of the
quasiparticle weight have to be consistently understood in a
nonperturbative framework.

As stated above, the polarization tensor does not show any
divergence with respect to the bandwidth cutoff $E_c$ and, at
the one-loop level, it is given by
\begin{equation}
i \Pi ({\bf k}, \omega_k ) = i \frac{e^2}{8} \frac{ {\bf
k}^2} {\sqrt{v_F^2 {\bf k}^2 - \omega_k^2 }}
\end{equation}
The dressed propagator of the interaction in the RPA is given by
\begin{equation}
\langle \phi ({\bf k}, \omega_k) \phi (-{\bf k}, -\omega_k)
\rangle = \frac{-i}{ 2 |{\bf k}| + \frac{e^2}{8}
 \frac{ {\bf k}^2}{\sqrt{v_F^2 {\bf k}^2 - \omega_k^2 }} }
\label{dress}
\end{equation}
By using the dressed propagator (\ref{dress}) in the computation
of the self-energy one is able to perform a partial sum of
perturbation theory, in which one takes into account the set of
most singular diagrams with regard to the bare interaction $\sim
1/ |{\bf k}|$. In this kind of GW approximation and for the
model with conical dispersion relation, we have
\begin{equation}
i \Sigma ({\bf k}, \omega_k ) = i 2 e^2 \int \frac{d^2 p}{(2
\pi)^2}
\frac{d \omega}{2 \pi} \; \frac{ \omega_k - \omega + v_F
\mbox{\boldmath $\sigma \cdot $} ({\bf k} - {\bf p} ) }
{ v_F^2 ({\bf k} - {\bf p} )^2 - (\omega_k - \omega )^2 } \;
\frac{-i}{ 2 |{\bf p}| + \frac{e^2}{8} \frac{ {\bf
p}^2}{\sqrt{v_F^2 {\bf p}^2 - \omega^2 }} }
\label{self}
\end{equation}
The imaginary part of the self-energy coming from ({\ref{self})
has been computed elsewhere\cite{unc}, and it has been shown to have a
linear dependence on quasiparticle energy, consistent with the
measured quasiparticle lifetimes in graphite. In this paper we
are interested in the computation of the real part of $\Sigma
({\bf k}, \omega_k )$, which provides information about the
nontrivial scaling of the quasiparticle weight and the Fermi
velocity in the low-energy limit.

The terms linear in $\omega_k$ and ${\bf k}$ in the self-energy
(\ref{self}) display a logarithmic dependence on the
high-energy cutoff $E_c$. This can be determined in the
following way. Since we are interested in the real part of
$\Sigma $ we may perform the analytic continuation $\omega
\rightarrow i \: \overline{\omega}$. We end up with the
expression:
\begin{eqnarray}
i \Sigma ({\bf k}, \omega_k ) & = &  i \frac{e^2}{v_F} \int
\frac{d^2 p}{(2 \pi)^2} \frac{d \overline{\omega}}{2 \pi} \;
\frac{1}{|{\bf p}|} \;  \frac{ \overline{\omega}_k -
\overline{\omega} +  \mbox{\boldmath $\sigma \cdot $} (v_F
{\bf k} - {\bf p} ) } { (v_F {\bf k} - {\bf p} )^2 +
(\overline{\omega}_k - \overline{\omega} )^2 }  \nonumber   \\
&   &  -  i \frac{e^2}{v_F} \frac{e^2}{16 v_F} \int \frac{d^2
p}{(2 \pi)^2} \frac{d \overline{\omega}}{2 \pi} \; \frac{\sqrt{
{\bf p}^2 + \overline{\omega}^2 } - \frac{e^2}{16 v_F} |{\bf
p}| } { \alpha {\bf p}^2 + \overline{\omega}^2 } \;  \frac{
\overline{\omega}_k - \overline{\omega} +  \mbox{\boldmath
$\sigma \cdot $} (v_F {\bf k} - {\bf p} ) } { (v_F {\bf k} -
{\bf p} )^2 + (\overline{\omega}_k - \overline{\omega} )^2 }
\end{eqnarray}
where $\alpha \equiv 1 - (e^2/(16 v_F))^2$. A singularity at
$\alpha = 0$ appears in the above expression, whose role has to
be clarified since there is no sign of a particular feature at
$e^2/(16 v_F) = 1$ in the original expression. We will perform
the above integrals taking the value of $\alpha > 0$, but it
will become clear at the end that the results can be continued
smoothly to the strong coupling regime $e^2/(16 v_F) > 1$.

Upon integration of the frequency from $-\infty$ to $+\infty$
and placing the bandwidth cutoff $E_c$ in momentum space, $v_F |{\bf
p}| < E_c$, the coefficients of the logarithmically divergent
contributions can be computed in terms of elementary functions
of $g \equiv e^2/(16 v_F)$. The renormalization of the electron
propagator turns out to be given by
\begin{eqnarray}
\frac{1}{G}  & = &  \frac{1}{G_0} - \Sigma    \nonumber    \\
  & = &  \; Z^{-1} (\omega_k - v_F  \mbox{\boldmath $\sigma \cdot$}
{\bf k}) \nonumber                          \\
  &  &    - Z^{-1}
\omega_k  \frac{8}{\pi^2}  \left( g^2 + \left( 2 -
g^2 \right) \left( 1 - \frac{\arcsin g}{g \sqrt{1-g^2}} \right)
\right)   \log E_c - Z^{-1} \omega_k \frac{8}{\pi}
\frac{1}{g} \left( \frac{1 - g^2/2}{\sqrt{1-g^2}} - 1  \right)
\log E_c  \nonumber \\
   &  &  + Z^{-1} v_F  \mbox{\boldmath
$\sigma \cdot$} {\bf k} \frac{8}{\pi^2}  \left(
 1 - \frac{\sqrt{1-g^2}}{g} \arcsin g \right) \log
E_c -  Z^{-1} v_F  \mbox{\boldmath $\sigma \cdot$} {\bf k}
\frac{4}{\pi} \frac{1}{g} \left( 1 - \sqrt{1-g^2} \right) \log
E_c
\end{eqnarray}
where $Z^{1/2}$ represents the
scale of the bare electron field compared to that of the
cutoff-independent electron field
\begin{equation}
\Psi_{bare} (E_c) = Z^{1/2} \Psi
\end{equation}

In the RG approach, we require the cutoff-independence of the
renormalized Green function, since this object leads to
observable quantities in the quantum theory. For this purpose,
$Z$ and $v_F$ have to be understood as cutoff-dependent
effective parameters, that reflect the behavior of the quantum
theory as $E_c \rightarrow 0$ and more states are integrated out
from high-energy shells of the band. We get the RG flow
equations
\begin{eqnarray}
E_c \frac{d}{d E_c} \log Z (E_c )  & = & - \frac{8}{\pi^2}
\left( 2 + \frac{2-g^2}{g} \frac{\arccos g}{\sqrt{1-g^2}}
\right)  +  \frac{8}{\pi} \frac{1}{g}  \label{zflow}        \\
E_c \frac{d}{d E_c} v_F (E_c )  & = & - \frac{8}{\pi^2} v_F
\left( 1 +  \frac{\arccos g}{g \sqrt{1-g^2}}
\right)  +  \frac{4}{\pi} v_F \frac{1}{g} \label{vflow}
\end{eqnarray}

Given that the electron charge $e$ is not renormalized, we may
write down the flow equation for the effective coupling constant
$g = e^2/(16 v_F)$
\begin{equation}
E_c \frac{d}{d E_c} g (E_c )  =  \frac{8}{\pi^2}
\left( g +  \frac{\arccos g}{\sqrt{1-g^2}}
\right)  -  \frac{4}{\pi} 
\label{gflow}
\end{equation}
In the weak coupling regime, one may check that the
renormalization of both the Fermi velocity and the electron
wavefunction takes place in the expected direction. The
quasiparticle weight $Z$ at small $g$ is smaller than the bare
value measured before integration of high-energy modes.
The Fermi velocity $v_F$ flows to higher values in the
infrared, and the density of states around the Fermi energy decreases,
as a consequence of screening effects.
This ensures the consistency of the weak coupling
phase, where the results of perturbation theory become increasingly
reliable in the low-energy limit. However, the most important
point concerns the possible existence of a different phase at
large values of $g$. In this respect, the flow equations
(\ref{zflow}) and (\ref{gflow}) can be analytically continued to
values $g > 1$, by simple use of the formula $\arccos z = i \log
(z - i \sqrt{1-z^2})$. The flows of the coupling constant and
the electron wavefunction are then differentiable across $g =
1$, which shows that the apparent singularity at this point has
no physical meaning.

The right-hand-side of Eq. (\ref{gflow}) is a monotonous 
function of $g$, taking into account the mentioned analytic
continuation. This means that
there is no phase different from that of the perturbative
regime, and that the strong coupling regime is connected to it
through RG transformations. The
RG flow represented in Fig. \ref{one} shows that the
perturbative regime is attained at low-energies, starting from
fairly large bare values of the coupling constant.

The present analysis is relevant to the phenomenology of the
graphite layers. It shows that, even in such a system with
unconventional quasiparticle lifetimes, the low-energy behavior
is governed by a fixed-point which can be described as a
Fermi liquid, as $Z$ tends to a constant in the infrared, 
unlike for the line of 
nontrivial fixed-points which characterize Luttinger liquids.
The low-energy scaling of the quasiparticle weight is represented
in Fig. \ref{two}.
Hence we can assert that, although the
quasiparticles decay according to the experiment with a rate
proportional to their energy, rather than to their energy
square, the notion of well-defined quasiparticles with a finite
weight at the Fermi level still holds in the system.

Our study stresses the anomalous screening properties
of the Coulomb interaction in low-dimensional systems.
This fact has been also put forward recently in a different
framework, pointing out that in one and two-dimensional systems
the screening of the long-range interactions goes in the
direction of reducing the electron correlations\cite{sw}.  
In the RG
approach we see how this effect arises under the form of a
renormalization of $v_F$. It is remarkable that such
nontrivial scaling of the Fermi velocity in the infrared
operates in the two-dimensional as well as in the
one-dimensional system with Coulomb interaction\cite{gb}, 
and it seems to be also present in systems with magnetic 
interactions\cite{wilczek,chakrav}.

The analysis presented here is relevant to the controversy about the
existence of systems with Fermi liquid behavior in two
dimensions. Our approach is completely
rigorous in the limit of a very large number $N$ of
electron flavors, where the bubble summation
implicit in the RPA becomes exact. 
The leading order in the $1/N$ expansion provides, on the other
hand, the fixed-point and the exact anomalous dimension of
the electron field in the Luttinger model\cite{metz,gb}. 
Therefore, this approximation should suffice to describe such
kind of nontrivial fixed-point,
in case that it were present in the model.

To summarize, semimetals described by the two-dimensional Dirac 
equation, such as a graphite layer, show 
significant differences with respect to the properties
of standard Fermi liquids with  (screened) Coulomb interactions.
In the present problem, the quasiparticle lifetime 
goes like $\sim \omega^{-1}$, while the enhancement 
of the Fermi velocity implies the vanishing of the effective
coupling in the infrared. 
We expect the same behavior in three-dimensional zero-gap semiconductors,
which are also described by an effective Dirac equation.

In the context of more general long-range interactions,
the system with a Dirac sea behaves also differently with
respect to the conventional Fermi sea, as in the former
an interaction $V(q) \sim 1/|q|^{1 + \epsilon}$
already departs from the Fermi liquid universality class for
$\epsilon > 0$. In our RG framework the interaction
becomes relevant no matter how small $\epsilon$ may be, and
a nontrivial fixed-point can be found away from the origin
within the $\epsilon$ expansion.

The logarithmic behavior that we have found affects also
some thermodynamic quantities like the specific heat 
or the susceptibility,
which pick up logarithmic corrections as $T \rightarrow 0$.
These are the signature of the marginal character of the
interaction, which has in our model the precise degree
of singularity to separate regimes with Fermi liquid ($\epsilon 
< 0$) and non-Fermi liquid behavior ($\epsilon > 0$).

\begin{figure}
\caption{Flow of the coupling constant for different bare values.}
\label{one}
\end{figure}

\begin{figure}
\caption{Wavefunction renormalization for bare
coupling constant $g = 5$ (thick line) and $g = 1$ (thin line).}
\label{two}
\end{figure}

\newpage

\par
\centering
\epsfbox[0 800 327 1127]{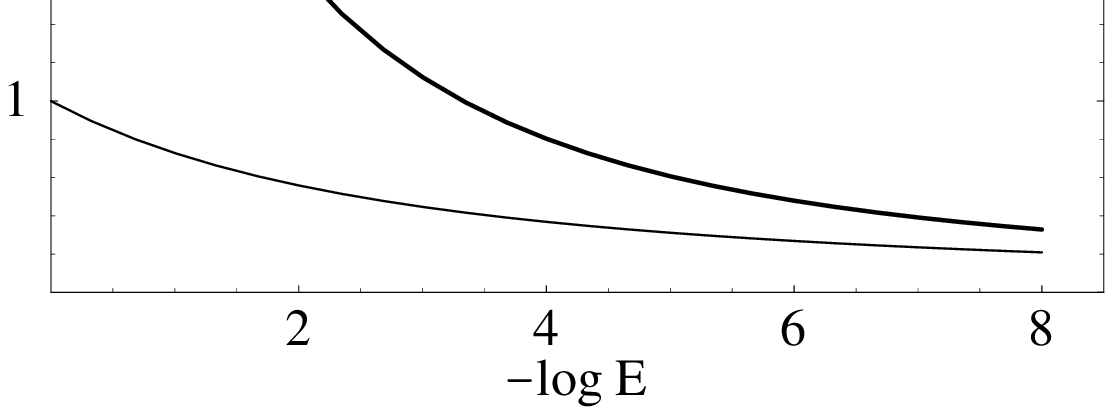}
\par

\newpage

\par
\centering
\epsfbox[0 800 312 1112]{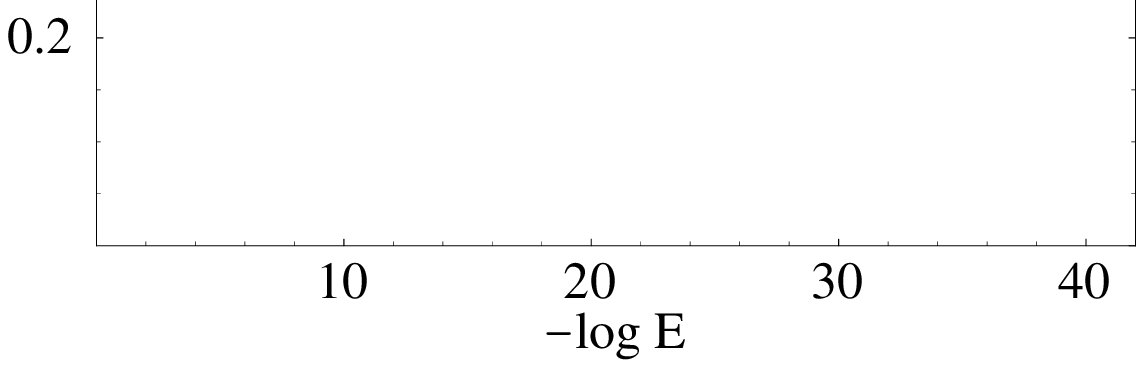}
\par


\begin{references}
\bibitem{foot}
We stick from now on to one of the Fermi points, which provides
all the relevant physics since the Coulomb interaction becomes
singular at small momentum transfer.
\end{references}
\end{document}